%% file: ptt-opp-dual.tex
\documentclass[submission,copyright,creativecommons]{eptcs}
\providecommand{\event}{LSFA 2021} 
\usepackage{underscore}           

\usepackage[british]{babel}
\usepackage{stmaryrd}
\usepackage{mathrsfs}
\usepackage{csquotes}
\usepackage{hyperref}
\usepackage{amsthm}
\usepackage{booktabs} 
\usepackage{enumitem}

\input{commands}

\renewcommand{\sneg}{\mathop{\neg}}

\newcommand{\equivf}{\Leftrightarrow}
\newcommand{\sequivf}{\mathbin{\Leftrightarrow_{\msf{s}}}}

\newcommand{\ulc}{\mathop{\circ}}
\newcommand{\dulc}{\mathop{\bullet}}

\renewcommand{\blc}{\mathbin{\scriptstyle \square}}
\newcommand{\dblc}{\mathbin{\scriptstyle \blacksquare}}

\newcommand{\ints}{\mathbin{\triangleq}}

\newcommand{\intsu}[1]{\mathbin{\triangleq_{#1}}}

\newcommand{\fs}[1]{\msf{#1}}
\newcommand{\itt}{\fs{ITT}}
\newcommand{\ptt}{\fs{PTT_0}}
\newcommand{\eptt}{\fs{PTT_1}}
\newcommand{\pls}{\fs{PL^S_0}}
\newcommand{\epls}{\fs{PL^S_1}}
\newcommand{\twoint}{\fs{2Int}}

\newcommand{\conc}[1]{\mr{C}_{#1}}

\newcommand{\invo}{\mathop{*}}

\newcommand{\refutationLine}{\dottedLine}

\newcommand{\cass}[1]{\cdot[#1]\cdot}

\newtheorem{thm}{Theorem}[section]

\theoremstyle{definition}
\newtheorem{defn}[thm]{Definition}

\title{About Opposition and Duality in Paraconsistent Type Theory}
\author{%
  Juan C. Agudelo-Agudelo%
  \institute{Universidad de Antioquia\\Medell\'in, Colombia}
  \email{juan.agudelo9@udea.edu.co}%
  \and%
  Andr\'es Sicard-Ram\'irez%
  \institute{Universidad EAFIT\\ Medell\'in, Colombia}%
  \email{asr@eafit.edu.co}%
}%
\def\titlerunning{About Opposition and Duality in Type Theory}
\def\authorrunning{J.~C. Agudelo-Agudelo \& A. Sicard-Ram\'irez}

\begin{document}
\maketitle

\begin{abstract}
  A paraconsistent type theory (an extension of a fragment of intuitionistic
  type theory by adding \enquote{opposite types}) is here extended by adding
  co-function types. It is shown that, in the extended paraconsistent type
  system, the opposite type constructor can be viewed as an involution
  operation that transforms each type into its dual type. Moreover, intuitive
  interpretations of opposite and co-function types under different
  interpretations of types are discussed.
\end{abstract}

\section*{Introduction}
\pdfbookmark[1]{Introduction}{sec:introduction}

A \emph{paraconsistent type theory}, called here~$\ptt$, was
introduced by the authors by extending~$\itt$ (the
$\to, \times, +, \Pi, \Sigma$ intensional fragment of intuitionistic
type theory, as presented in~\cite{martin-lof-1984}, with two
universes~$\uzero$ and~$\uone$) with the addition
of \emph{opposite types} in~\cite{agudelo-sicard-2021}. In~$\ptt$, for
each type~$A$, there is an opposite type~$\op{A}$. The introduction
and elimination rules for opposite types were defined for each one of
the type constructors (including the opposite type constructor itself), and
such rules were based on the rules for \emph{constructible falsity}
\cite{nelson-1949,almukdad-nelson-1984,lopez-escobar-1972}. A
propositions-as-types correspondence between~$\pls$ (a many-sorted
version of the \emph{refutability calculus}--introduced by
L\'opez-Escobar in~\cite{lopez-escobar-1972}--presented in natural
deduction style) and~$\ptt$ was proven.\footnote{The logic~$\pls$ is denoted
in~\cite{agudelo-sicard-2021} by~$\fs{PL_S}$.} Under such propositions-as-types
correspondence, the opposite type constructor corresponds to negation
in~$\pls$, the correspondence for the other type constructors are the same than
for~$\itt$ with respect to intuitionistic logic.

Differently from how it is done in intuitionist type theory, where
negation is formalised by means of the function type and the empty
type, negation in~$\ptt$ is formalised by the primitive opposite type
constructor, without need of the empty type. Under such formalisation,
an inhabitant of a type~$\op{A}$ can be understood as a proof term
for~$\sneg A$ or as a \enquote{refutation term} for~$A$, and proofs
and refutations are treated in a symmetric and constructive way.

As $\pls$ is a paraconsistent logic (i.e.~for some set of formulae
$\Delta$ and some formulae $A$ and $B$ we have that
$\Delta \vdash_{\pls} A$, $\Delta \vdash_{\pls} \sneg A$ and
$\Delta \nvdash_{\pls} B$), the propositions-as-types correspondence
with~$\ptt$ leads to the existence of logically contradictory but not
trivial contexts (i.e.~contexts~$\Gamma$ for which there exist a
type~$A$ and terms~$t$ and ~$s$ such that ~$\Gamma \vdash_{\ptt} t: A$
and $\Gamma \vdash_{\ptt} r:\op{A}$, but~$\Gamma \nvdash_{\ptt} s: B$,
for some type~$B$ and every term~$s$).  Because of that, $\ptt$~is
considered a \enquote{paraconsistent type theory} (for details
see~\cite{agudelo-sicard-2021}).


We wrote the following observation in the section of concluding
remarks and future work of~\cite{agudelo-sicard-2021}:
\begin{displayquote}
  The opposite type constructor can be viewed as an operation that
  transforms types into their \enquote{duals}. Under a logical (or
  propositions-as-types) interpretation, the duality is between truth
  and falsity of propositions (since the habitation of a type~$\op{A}$,
  interpreted as a proposition, can be understood as~$A$ is false). If
  we interpret types as problems (or tasks), the duality is between
  solvability and unsolvability of the respective problems. A more
  difficult understanding of the duality is when types are interpreted
  as sets. However, also under the set interpretation, opposite types
  can be viewed as an operation that establishes a duality between some
  set operations, although it is difficult to understand in what sense
  this duality occurs.
\end{displayquote}

Despite such observation, the notion of duality in~$\ptt$ was neither
rigorously defined nor discussed in depth
in~\cite{agudelo-sicard-2021} and this is just the aim of this
article.

In~$\pls$, the logical constants $\impl, \conj, \disj, \forall$ and $\exists$
have the same introduction and elimination rules than in intuitionistic logic,
but negation is a primitive logical constant, not definable in terms of
the other logical constants, and its behaviour is established by defining
introduction and elimination rules for the negation of each one of the logical
constants (including negation itself). Moreover, equivalence $\equivf$ can be
defined in the usual way (i.e.~$A \equivf B \eqdef (A\impl B) \conj (B\impl
A)$), but substitution by equivalents fails. However, a \emph{strong
equivalence} $\sequivf$ can be defined by $A \sequivf B \eqdef (A \equivf B)
\conj (\sneg A \equivf \sneg B)$ and substitution by strong equivalents is
valid. The following strong equivalences are derivable in~$\pls$ (where
superscripts on quantified variables denote their sorts):
\begin{align*}
  \sneg (A \conj B) &\sequivf \sneg A \disj \sneg B, & \sneg \foralls{x}{s}{A} &\sequivf \existss{x}{s}{\sneg A},
  \\
  \sneg (A \disj B) &\sequivf \sneg A \conj \sneg B, &\sneg \existss{x}{s}{A} &\sequivf \foralls{x}{s}{\sneg A},
                    & \sneg \sneg A &\sequivf  A.
\end{align*}
These strong equivalences show that in~$\pls$ the negation is an
involutive operation under which there is duality between $\conj$ and
$\disj$, and also between $\forall$ and $\exists$. It is important to
highlight here that, although in~$\pls$ the De Morgan's laws, the
\enquote{classical} equivalences for the negation of quantifiers and
the double negation law are valid, this does not make this logic
collapse to classical logic. There are many formulae valid in classic
logic which are not valid in~$\pls$ including ${A \disj \sneg A}$,
${A \impl (\sneg A \impl B)}$ and ${\sneg (A \conj \sneg A)}$. The
non-validity of this formulae can be proven by using the Kripke style
semantics provided for L\'opez-Escobar's refutability calculus
in~\cite{lopez-escobar-1972} (which can be naturally adapted
to~$\pls$).

For implication,
although~$\vdash_{\pls} \sneg (A \impl B) \equivf (A \conj \sneg B)$,
we have
that~$\nvdash_{\pls} \sneg \sneg (A \impl B) \equivf \sneg (A \conj
\sneg B)$,
thus~$\nvdash_{\pls} \sneg (A \impl B) \sequivf (A \conj \sneg
B)$. Consequently, if we require that the dual of a formula be
strongly equivalent to its negation, so that the negation of the
formula and its dual be inter-substitutable (which seems to be a
reasonable requirement), then $A \conj \sneg B$ cannot be considered a
dual of~$A \impl B$ in~$\pls$. Moreover, under the previous
requirements, there is no dual logical constant to~$\impl$ in~$\pls$,
but in order to every logical constant have its dual, the logic~$\pls$
is extended in Section~\ref{sec-ptt-to-eptt} by including a
co-implication logical constant whose rules are taking
from~\cite{wansing-2016a} (where a kind of bi-intuitionistic logic,
named $\twoint$, is defined by dualising the natural deduction rules
of intuitionistic propositional logic). A rigorous definition of dual
logical constants is given in Definition~\ref{def-duality-log}, and
under such definition the dualities previously mentioned are
justified.

As the introduction and elimination rules for opposite types in~$\ptt$
are based on the rules of the (constructible) negation of~$\pls$, the
opposite type constructor works as an involutive operation under which
there is duality between~$\times$ and~$+$, and between $\Pi$ and
$\Sigma$. As in~$\pls$ there is no dual type constructor to~$\to$,
this type system is extended in Section~\ref{sec-ptt-to-eptt} by
adding a new constructor of \enquote{co-function types}, whose rules
are based on the rules for co-implication in the extension of
$\pls$. The system obtained is called~$\eptt$ and one of its
characteristics is that every type constructor have a dual, which
allows a total symmetric formalisation of proofs and refutations.  A
rigorous definition of dual type constructors is given in
Definition~\ref{def-duality-tt}, and under such definition the
dualities previously mentioned are justified.

This article is structured as follows: The type system~$\ptt$ is
described in Section~\ref{sec-ptt}. In Section~\ref{sec-ptt-to-eptt},
$\ptt$ is extended to~$\eptt$. In Section~\ref{sec-dual-eptt}, it is
shown that the opposite type constructor in~$\eptt$ can be viewed as
an involution operation that transforms each type into its dual type,
and it is stated a \enquote{principle of duality} in~$\eptt$. Finally,
some observations about the intuitive interpretation of opposite and
co-function types in~$\eptt$ are presented in
Section~\ref{sec-opp-cof-int}.  With respect to the interpretation of types as
sets, we only provide a vague description of how opposite types can be
generally understood. Although a reviewer reasonably suggested that duality is
not set-theoretic and we should just forget about this interpretation, we do
not want to abandon this possibility for the moment, and we think that the
vague idea that we present may shed some light in further developments of this
possible interpretation.

\section{The type system~\texorpdfstring{$\ptt$}{PTT}}\label{sec-ptt}

We shall give a brief description of system~$\ptt$, for further
details see~\cite{agudelo-sicard-2021}. As we mentioned in the
introduction, $\ptt$ is introduced by extending~$\itt$ (the
$\to, \times, +, \Pi, \Sigma$ intensional fragment of intuitionistic
type theory with two universes $\uzero$ and $\uone$) with the addition
of \emph{opposite types}. The formation rule for opposite types
in~$\ptt$ is:
\begin{equation*}
  \pttopf.
\end{equation*}
Taking into consideration that the introduction and elimination rules
for the constructors~$\to, \times, +, \Pi,\Sigma$ in~$\itt$ are
well-known, only the introduction and elimination rules for opposite
types in~$\ptt$ are described in Table~\ref{tab-rules-opp} (where the
term constructors are the same of~$\itt$ as presented
in~\cite{martin-lof-1984}).

\begin{table}[!ht]
  \renewcommand{\arraystretch}{2}
  \center
  \begin{tabular}{@{}|c|c|@{}}
    \hline
    \pttopopi & \pttopope \rule[-1em]{0pt}{3em}%
    \\
    \hline
    \pttopfunci & \pttopfuncel \quad \pttopfuncer \rule[-1.5em]{0pt}{3em}%
    \\
    \hline
    \parbox[c]{2.25cm}{%
      \medskip%
      \pttopprodil%
      \medskip%
      \medskip%
      \pttopprodir%
      \medskip%
    } & \pttopprode
    \\
    \hline
    \pttopdijui & \pttopdijuel \quad \pttopdijuer \rule[-1.5em]{0pt}{3em}%
    \\
    \hline
    \pttoppii & \rule{0pt}{4em}%
    \pttoppie%
    \rule[-4em]{0pt}{4em}%
    \\
    \hline
    \rule{0pt}{4em}%
    \pttopsigi%
    \rule[-4em]{0pt}{3em}%
    & \pttopsige
    \\
    \hline
  \end{tabular}
  \caption{Introduction and elimination rules for opposite types
    in~$\ptt$.}\label{tab-rules-opp}
\end{table}

The choice of terms for opposite types in~$\ptt$ is based on
L\'opez-Escobar's extension of the so-called BHK-interpretation of
intuitionistic logic~\cite{lopez-escobar-1972}, where the definition
of what constitutes a construction to refute a formula is made using
the same objects that constitute proofs of formulae in the
BHK-interpretation, taking advantage of the duality between logical
constants and the assumption that a construction~$c$ refutes a
formula~$A$ iff~$c$ proves~$\sneg A$. For instance, a construction~$c$
that refutes a formula $A \disj B$ is a pair ${c = (a, b)}$, where~$a$
and~$b$ are constructions that refute~$A$ and~$B$, respectively (in an
analogous and dual way as a construction~$c$ that proves a formula
$A \conj B$ is a pair ${c = (a, b)}$, where~$a$ and~$b$ are
constructions that prove~$A$ and~$B$, respectively). Consequently, the
term constructors for opposite types in~$\ptt$ already exists
in~$\itt$ and their computation (or equality) rules in~$\ptt$ are
defined just in the same way as in~$\itt$. \footnote{This choice of
  terms leads to non-uniqueness of types in~$\ptt$, but this does not
  seem to be an inconvenience in this type
  system~\cite[\S~6]{agudelo-sicard-2021}.  However, several causes of
  the non-uniqueness of types in~$\ptt$ are avoided by the inclusion
  of the equality rules for opposite types in
  Table~\ref{tab-eq-rules-opp}.}

Universes~$\uzero$ and~$\uone$, with $\uzero :\uone$, are introduced
in~$\ptt$ in order to prove a propositions-as-types correspondence
between this type system and the logic~$\pls$. Universe~$\uzero$ is
closed under~$\to, \times, +, \Pi,\Sigma$ and opposite types,
while~$\uone$ is closed only under~$\to$. The only aim of~$\uone$ is
to allow the creation of types corresponding to predicates, for which
it is enough to have in~$\uone$ the type constructor~$\to$. Thus the
restriction of~$\uone$ to be closed only under~$\to$ is to facilitate
the proof of the propositions-as-types correspondence. Consequently,
the idea of opposite types is developed in~$\ptt$ only in
universe~$\uzero$, although it can be naturally extended to other
universes. Using widespread terminology, types in~$\uzero$ are called
\emph{small types}.

\section{Extending~\texorpdfstring{$\ptt$}{PTT} with
co-function types}\label{sec-ptt-to-eptt}

The familiar introduction and elimination rules for intuitionistic
propositional logic are dualised by employing a primitive notion of
\emph{dual proof} by Wansing in~\cite{wansing-2016a}. Here, we prefer
to use \emph{refutation} instead of dual proof. By using single-line
rules for proofs and dotted-line rules for refutations, and
using~$\impl$ and~$\coimpl$ as the logical constants for implication
and co-implication, respectively,\footnote{In~\cite{wansing-2016a},
  the logical constants for implication and co-implication
  are~$\impli$ and~$\coimpli$, respectively, but we shall reserve
  these symbols for function and co-function types. Moreover, we
  changed the notation for refutations and counter-assumptions, in
  order to avoid confusions with our notation for opposite types.} the
dual rules for~$\impl$ are:
\begin{align*}
  \ndtilimpli{} \quad \leadsto \quad \ndtilimplid{,} & & \ndtilimple{}
                                                         \quad \leadsto
                                                         \quad \ndtilimpled{,}
\end{align*}
where~$\leadsto$ means \enquote{dualises to}, $[A]$~denotes the
cancellation of the assumption~$A$ in the conclusion and~$\cass{A}$
denotes the cancellation of counter-assumption~$A$ (or the assumption
of the falsity of~$A$) in the conclusion.  Formula ${B \coimpl A}$ may
be read as \enquote{$A$ co-implies~$B$} or as \enquote{$B$
  excludes~$A$} (see~\cite[Footnote~2]{wansing-2008}).

As the process of dualisation does not induce rules for the
falsification of implications nor for the verification of
co-implications, these rules are defined in~\cite{wansing-2016a} by
taking an orthodox stance as follows:\footnote{We changed the order of
  formulae in the co-implication rules for a better understanding of
  these rules as duals of implication rules.}
\begin{align*}
  \ndtilnimpli{,} & & \ndtilnimplel{,} \qquad \ndtilnimpler{,}
  \\
  \\
  \ndtilncoimpli{,} & & \ndtilncoimpler{,} \qquad \ndtilncoimplel{.}
\end{align*}

As in~$\pls$ negation represents falsity, and a proof of~$\sneg A$ can
be also understood as a refutation of~$A$, we can extend~$\pls$ with
co-implication and define the introduction and elimination rules
without using dotted-lined rules for refutations and dotted brackets
for counter-assumptions as follows.


\begin{align*}
  \nlcoimpli{,} & & \nlcoimplel{,} \qquad \nlcoimpler{,}
  \\
  \\
  \nlncoimpli{,} & & \nlncoimple{.}
\end{align*}

The extension of~$\pls$ with co-implication will be called~$\epls$.\footnote{In
  \cite{wansing-2008}, two ways of formalising co-implication are presented. In
  one way, co-implication is strongly equivalent to negated implication (i.e. $A
  \coimpl B \sequivf \sneg(A \to B)$). In the other way, co-implication is
  strongly equivalent to negated contrapose implication (i.e. $A  \coimpl B
  \sequivf \sneg(\sneg B \to \sneg A)$). In $\pls$, contraposition is not valid
  and the two ways of formalising co-implication are not equivalent. Under our
  formalisation of co-implication in~$\epls$, it is possible to prove that $A
  \coimpl B \sequivf \sneg(\sneg B \to \sneg A)$, which shows that the
  convincing
  process of dualisation in~\cite{wansing-2016a} leads co-implication to behave
  as negated contraposed implication, instead of behaving as negated
  implication.}
\footnote{From Negri and von Plato's generalisation of the \emph{inversion
principle}, which states that \enquote{whatever follows from the direct grounds
for deriving a proposition must follow from that
proposition}~\cite[p.~6]{negri-vonplato-2001}, general elimination
rules are uniquely determined by each introduction
rule. For the case of co-implication, the
general elimination rule would be:
\begin{equation*}
  \AxiomC{$B \coimpl A$}%
  \AxiomC{$[\sneg A, B]$}%
  \noLine%
  \UnaryInfC{$\vdots$}%
  \noLine%
  \UnaryInfC{}%
  \noLine%
  \UnaryInfC{$C$}%
  \BinaryInfC{$C$}%
  \DisplayProof%
\end{equation*}

This general elimination rule differs from the two standard elimination rules
presented above. However, as in the case of intuitionistic logic, if we change
in~$\epls$ the standard elimination rules by the general elimination rules, the
system obtained is equivalent with respect to deductibility. In the case of
intuitionistic logic, the difference emerges in the correspondence with sequent
calculus: Normal derivations in the natural deduction system with general
elimination rules can be isomorphically translated into cut-free derivations in
the sequent calculus with independent contexts, which is not possible with the
standard elimination rules~\cite[Ch.~8]{negri-vonplato-2001}. As we are not
interested in establishing a correspondence of the natural deduction system
for~$\epls$ with a sequent calculus, we choose the standard elimination rules
for co-implication (and all the other logical constants).

In comparison with linear logic, which contains a fully involutive
negation and can be seen \enquote{as a bold attempt to reconcile the
  beauty and symmetry of the systems for classical logic with the
  quest for constructive proofs that had led to intuitionistic
  logic}~\cite{dicosmo-miller-2019}, the same can be said
for~$\epls$. However, linear logic is obtained by eliminating the
contraction and weakening rules of a sequent calculus for classical
logic, allowing the formalisation of two different versions of each
logical constant: An \emph{additive} version (where the contexts of
the premises are the same) and a \emph{multiplicative} version (where
the contexts of the premises can be different); while the symmetry
in~$\epls$ is obtained by the formalisation of a primitive
constructive negation whose rules are based on the understanding of
negation as falsity, and on the notion of refutation which is dual to
the notion of proof. The distinction between additive and
multiplicative logical constants in~$\epls$ is at least not
evident. Although a reviewer pointed us out that the unique general
elimination rule for co-implication leads to a multiplicative version
of the connective, while the two standard elimination rules leads to
an additive version of the connective, the connection between
multiplicativity/additivity (which are concepts usually defined when
working with sequent calculus) and
general-elimination-rules/standard-elimination-rules (which are
concepts usually defined in natural deduction systems presented in
standard format, that is, where the rules are presented without
entailment relations nor contexts) is not clear for us. In
\cite{negri-2002a}, where a natural deduction system (in standard
format) for intuitionistic linear logic is proposed, all elimination
rules (for additive and multiplicative logical constants) are
general. The additivity of logical constants is formalised by adding
labels to the assumptions in context-sharing rules.}

Now, we shall extend~$\ptt$ by adding co-function types, whose rules
will be based on the co-implication rules. Symbol~$\cofunc$ will be
used to denote the co-function type constructor and~${B \cofunc A}$
may be read as \enquote{the type of co-functions from~$A$ to~$B$}. The
formation rule for co-function types is:
\begin{equation*}
  \pttcofuncf{.}
\end{equation*}
The introduction and elimination rules for co-function types and their
opposites are presented in Table~\ref{tab-rules-cofunc}.

\begin{table}[!ht]
  \center
  \begin{tabular}{@{}|c|c|@{}}
    \hline
    \pttcofunci{} & \pttcofuncel{} \quad \pttcofuncer{} \rule[-1.25em]{0pt}{3em}
    \\
    \hline
    \rule{0pt}{4.0em} \pttopcofunci{}  \rule[-3.5em]{0pt}{3em} & \pttopcofunce{}
    \\
    \hline
  \end{tabular}
  \caption{Introduction and elimination rules for
  co-function types and their opposites.}\label{tab-rules-cofunc}
\end{table}

As the terms used in the rules for co-function types already exits
in~$\itt$, their computation rules are defined in the same way as
in~$\itt$.  The extension of~$\ptt$ with co-function types will be
called~$\eptt$.

With some laborious but not difficult work, the propositions-as-types
correspondence between~$\pls$ and~$\ptt$ can be extended to obtain a
propositions-as-types correspondence between~$\epls$
and~$\eptt$. Under such extended correspondence, the co-function type
constructor will correspond to co-implication.

\section{Duality in~\texorpdfstring{$\eptt$}{PTT'}}\label{sec-dual-eptt}

As it is pointed out in~\cite[p. 187]{gowers-barrow-leader-2008}:

\begin{displayquote}
  Duality is an important general theme that has manifestations in
  almost every area of mathematics. Over and over again, it turns out
  that one can associate with a given mathematical object a related,
  \enquote{dual} object that helps one to understand the properties of
  the object one started with. Despite the importance of duality in
  mathematics, there is no single definition that covers all instances
  of the phenomenon.
\end{displayquote}

Although there is not a general definition of duality in mathematics,
we shall take the description in~\cite{wiki-duality-math-2021}, which
is clear and general enough for our purposes here.

\begin{displayquote}
  In mathematics, a \emph{duality} translates concepts, theorems or
  mathematical structures into other concepts, theorems or structures,
  in a one-to-one fashion, often (but not always) by means of an
  involution operation: if the dual of~$A$ is~$B$, then the dual
  of~$B$ is~$A$. Such involutions sometimes have fixed points, so that
  the dual of~$A$ is~$A$ itself.
\end{displayquote}

Before explaining duality in~$\eptt$, we shall provide a rigorous definition of
duality between logical constants in a logical system, and based on such
definition we shall state the dualities between logical constants in~$\epls$.

\begin{defn}\label{def-duality-log}
  Let~$\fs{L}$ be a logical system with consequence
  relation~$\Vdash_{\fs{L}}$:
  \begin{enumerate}[label=(\roman*)]
  \item Two formulae~$A$ and~$B$ are \emph{inter-substitutable
      in~$\fs{L}$}, which will be denoted by~$A \ints B$, if for every
      formula~$C$ of~$\fs{L}$, when~$C'$ is the result of substituting some
      occurrences of~$A$ by~$B$ (or vice versa) in $C$, then~$C \Vdash_{\fs{L}}
      C'$ and~$C' \Vdash_{\fs{L}} C$.

  \item A unary logical constant~$\invo$ is an \emph{involution in
      $\fs{L}$} if~$A \ints \invo (\invo A)$, for every formula~$A$ of~$\fs{L}$.

  \item Two unary logical constants~$\ulc$ and~$\dulc$ are \emph{dual
      in $\fs{L}$}, under an involution~$\invo$, if~$\invo (\ulc A) \ints
      \dulc(\invo A)$ and~$\invo(\dulc A) \ints \ulc(\invo A)$, for
    every formula~$A$ of~$\fs{L}$.

  \item Two binary logical constants~$\blc$ and~$\dblc$ are \emph{dual
      in $\fs{L}$}, under an involution~$\invo$, if
    ${\invo (A \blc B)} \ints {\invo A \dblc \invo B}$ and ${\invo (A \dblc B)}
    \ints {\invo A \blc \invo B}$, or if ${\invo (A \blc B)} \ints {\invo B
    \dblc \invo A}$ and ${\invo (B \dblc A)} \ints {\invo A \blc \invo B}$, for
    every pair of formulae~$A$ and~$B$ of~$\fs{L}$.
  \end{enumerate}
\end{defn}

Quantifiers are considered unary logical constants, and the binding
variables will be considered parameters that are part of the logical
constant. For instance, $\forall x$ will be considered a logical
constant, parametrised by $x$. When we say that quantifiers~$\forall$
and~$\exists$ are dual, we mean that they are dual under every
parameter (or binding variable)~$x$.

\begin{thm}
  In~$\epls$:
  \begin{enumerate}[label=(\roman*)]
  \item If $\vdash_{\epls} A \sequivf B$, then $A \ints B$.

  \item $\sneg$ is an involution.

  \item $\conj$ and~$\disj$ are dual logical constants under~$\sneg$.

  \item $\impl$ and~$\coimpl$ are dual logical constants
    under~$\sneg$.

  \item $\forall$ and~$\exists$ are dual logical constants
    under~$\sneg$.
  \end{enumerate}
\end{thm}

Now, we shall explain why the opposite type constructor in~$\eptt$ can
be viewed as an involution operation that transforms types into their
dual types. Firstly, we provide a rigorous definition of duality
between type constructors in a type theory.

\begin{defn}\label{def-duality-tt}
  Let~$\fs{T}$ be a type system with consequence
  relation~$\Vdash_{\fs{T}}$, and let~$\univ$ be a universe
  of~$\fs{T}$:
  \begin{enumerate}[label=(\roman*)]
  \item Two types~$A$ and~$B$ are \emph{inter-substitutable in~$\univ$}, which
  will be denoted by~$A \intsu{\univ} B$, if $A$ and $B$ are in $\univ$ and for
  every type~$C$ in~$\univ$, when~$C'$ is the  result of substituting some
  occurrences of~$A$ by~$B$ (or
  vice versa) in~$C$, then~$x:C \Vdash_{\fs{T}} x:C'$ and~$x:C' \Vdash_{\fs{T}}
  x:C$, for every variable~$x$ that is not in~$C$.

  \item A unary type constructor~$\invo $ is an \emph{involution in
      $\univ$} if~$A \intsu{\univ} \invo (\invo A)$, for every type~$A$
      in~$\univ$.

  \item Two unary type constructors~$\ulc$ and~$\dulc$ are \emph{dual
      in $\univ$}, under an involution~$\invo$, if~$\invo(\ulc A) \intsu{\univ}
      \dulc (\invo A)$ and~$\invo (\dulc A) \intsu{\univ} \ulc (\invo A)$, for
    every type~$A$ in~$\univ$.

  \item Two binary type constructors~$\blc$ and~$\dblc$ are \emph{dual
      in $\univ$}, under an involution~$\invo$, if
    ${\invo (A \blc B)} \intsu{\univ} {\invo A \dblc \invo B}$ and ${\invo (A
    \dblc B)} \intsu{\univ} {\invo A \blc \invo B}$, or if ${\invo (A \blc B)}
    \intsu{\univ} {\invo B \dblc \invo A}$ and ${\invo (B \dblc A)}
    \intsu{\univ} {\invo A \blc \invo B}$, for every pair of types~$A$ and~$B$
    in~$\univ$.
  \end{enumerate}
\end{defn}

Similarly as quantifiers are considered unary logical constants, we
shall consider the type constructors~$\Pi$ and~$\Sigma$ unary type
constructors, and the binding variables and their types will be
considered parameters that are part of the type constructor. For
instance, $\Pi x:A$ will be considered a unary type constructor
parametrised by~$x$ and $A$. When we say that~$\Pi$ and~$\Sigma$ are
dual, we mean that they are dual under every pair of parameters~$x$
and~$A$.

In~\cite{agudelo-sicard-2021}, equivalence and strong equivalence
relations between types of~$\ptt$ are defined. We adapt these
definitions for~$\eptt$.

\begin{defn}
  Let~$A$ and~$B$ be two small types of~$\eptt$.
  \begin{enumerate}[label=(\roman*)]
    \item $A$ and~$B$ are \emph{equivalent in $\uzero$}, which will be denoted
    by $A \equivt{\uzero} B$, if for every context~$\Gamma$ we have that $\Gamma
    \vdash_{\eptt} t:A$ iff~$\Gamma \vdash_{\eptt} t:B$.

    \item $A$ and~$B$ are \emph{strongly equivalent}, which will be denoted by
    $A \sequivt{\uzero} B$, if $A \equivt{\uzero} B$ and~$\op{A}
    \equivt{\uzero} \op{B}$.
  \end{enumerate}
\end{defn}

Taking into account that for types $\op{A \to B}$ and
$A \times \op{B}$ apply the same introduction and elimination rules,
we could think that $\op{A \to B} \equivt{\uzero} A \times \op{B}$ is a direct
consequence of such fact. The following derivation shows that
$x:\op{A \to B} \vdash_{\eptt} \pair{\proj{1}{x}}{\proj{2}{x}}: A
\times \op{B}$:
\begin{equation*}
  \AxiomC{$x:\op{A \to B}$}%
  \UnaryInfC{$\proj{1}{x}:A$}%
  \AxiomC{$x:\op{A \to B}$}%
  \UnaryInfC{$\proj{2}{x}:\op{B}$}%
  \BinaryInfC{$\pair{\proj{1}{x}}{\proj{2}{x}}: A \times \op{B}$}%
  \DisplayProof%
\end{equation*}
However, it is necessary to include a conversion rule into~$\eptt$ in
order to make~$\pair{\proj{1}{x}}{\proj{2}{x}} = x$ and prove
that~$x:\op{A \to B} \vdash_{\eptt} x: A \times \op{B}$. Similarly,
for types $\op{A \times B}$ and $\op{A} + \op{B}$ apply the same
introduction and elimination rules, and the following derivation shows
that
$z:\op{A \times B} \vdash_{\eptt} \eltd{z}{x}{\injl}{y}{\injr} :
\op{A} + \op{B}$:
\begin{equation*}
  \AxiomC{$z:\op{A \times B}$}%
  \AxiomC{$(x:\op{A})$}%
  \UnaryInfC{$\inj{1}{x}:\op{A} + \op{B}$}%
  \AxiomC{$(y:\op{B})$}%
  \UnaryInfC{$\inj{2}{x}:\op{A} + \op{B}$}%
  \TrinaryInfC{$\eltd{z}{x}{\injl}{y}{\injr}:\op{A} + \op{B}$}%
  \DisplayProof%
\end{equation*}
However, it is necessary to include a conversion rule into~$\eptt$ in
order to make ${\eltd{z}{x}{\injl}{y}{\injr} = z}$ and prove
that~$z:\op{A \times B} \vdash_{\eptt} z: \op{A} + \op{B}$. Analogous
situations occur when trying to proving some other apparently evident
equivalences between types, which justifies the inclusion of the
conversion rules presented in Table~\ref{tab-eta-rules} into~$\eptt$,
where
$W \in \big\{ A \to B, \pit{x}{A}{B(x)}, \op{B \cofunc A},
\op{\sigt{x}{A}{B(x)}} \big\}$,
$X \in \big\{ A \times B, B \cofunc A, \sigt{x}{A}{B(x)}, \op{A + B},
\op{A \to B}, \op{\pit{x}{A}{B(x)}} \big\}$,
$Y \in \big\{ A + B, \op{A \times B} \big\}$ and
$Z \in \big\{ \sigt{x}{A}{B(x)}, \op{\pit{x}{A}{B(x)}} \big\}$. The
rules in the first row of the table are called \emph{eta rules} and
the ones in the second row are called \emph{co-eta
  rules}.\footnote{For an in-depth discussion of eta and co-eta rules
  in Martin-Löf type theory see~\cite{klev-2019}.}

\begin{table}[!ht]
  \centering
  \begin{tabular}{@{}|c|c|@{}}
    \hline
    \pttetaconvabst{W}{} & \pttetaconvpair{X}{} \rule[-1.25em]{0pt}{3em}
    \\
    \hline
    \rule{0pt}{1.8em} \pttcoetaconvseld{Y}{} & \pttcoetaconvsels{Z}{}
    \rule[-1.25em]{0pt}{3em}\\
    \hline
  \end{tabular}
  \caption{Eta and co-eta conversion rules.}\label{tab-eta-rules}
\end{table}

With the addition of the eta and co-eta conversion rules, the following strong
equivalences between types of~$\eptt$ can be proven.

\begin{thm}\label{thm-sequiv-eptt}
  In~$\eptt$, for every two small types~$A$ and~$B$, we have the
  following strong equivalences:
  \begin{align*}
    \op{A \to B} &\sequivt{\uzero} \op{B} \cofunc \op{A}, & \op{A
                                                            \times B}
    &\sequivt{\uzero} \op{A} + \op{B}, & \op{\pit{x}{A}{B}} &\sequivt{\uzero} \sigt{x}{A}{\op{B}},
    \\
    \op{B \cofunc A} &\sequivt{\uzero} \op{A} \to \op{B}, & \op{A + B}
    &\sequivt{\uzero} \op{A} \times \op{B}, & \op{\sigt{x}{A}{B}}
                                                            &\sequivt{\uzero}
                                                              \pit{x}{A}{\op{B}},
                     & \op{\op{A}} &\sequivt{\uzero} A.
  \end{align*}
\end{thm}
Under the propositions-as-types interpretation
in~\cite{agudelo-sicard-2021}, equivalence and strong equivalence
in~$\pls$ do not correspond, respectively, to equivalence and strong
equivalence in~$\ptt$. For instance, we have that
${\vdash_{\pls} (A \conj A) \equivf A}$ (and also
${\vdash_{\pls} (A \conj A) \sequivf A}$), but
${(A \times A) \nequivt{\uzero} A}$ (and consequently~${(A \times A)
\nsequivt{\uzero} A}$) in~$\ptt$. Moreover, while~$\equivf$ and~$\sequivf$ are
logical constants defined in the object language of~$\pls$, the
relations~$\equivt{\uzero}$ and~$\sequivt{\uzero}$ are defined on the
meta-language. The same differences occur if the
propositions-as-types interpretation is extended to~$\epls$
and~$\eptt$. In certain way, the equivalence relation defined for
types is more exigent that the equivalence defined for formulae,
demanding not only equivalence with respect to deductibility (or
inhabitation) but also demanding that their proof terms (or
inhabitants) are just the same.

In intuitionistic type theory (and other type theories) a conversion
relation between types is defined in order to make equivalent types be
equal under such conversion relation, thus ensuring uniqueness in type
assignment. This is not possible in~$\eptt$ (and neither in~$\ptt$), because in
these systems there are equivalent types that are not strongly
equivalent (for instance, $\op{A \to B} \equivt{\uzero} A \times \op{B}$ but
$\op{\op{A \to B}} \nequivt{\uzero} \op{A \times \op{B}}$). When two types
of~$\eptt$ are strongly equivalent they work as being the same type and
\enquote{substitution by strongly equivalent types is possible}, what
does not happen if they are only equivalent (for instance, we have
that $\op{A \to B} \equivt{\uzero} A \times \op{B}$, but
$\op{\op{A \to B}} \nequivt{\uzero} \op{A \times \op{B}}$, consequently
$\op{A \to B}$ and $A \times \op{B}$ are not inter-substitutable in~$\eptt$).

\begin{thm}
  Let~$A$ and~$B$ be small types of~$\eptt$.
  \begin{enumerate}[label=(\roman*)]
    \item If~$A\sequivt{\uzero} B$, then~$A \intsu{\uzero} B$.

    \item If~$A \intsu{\uzero} B$, then for every context~$\Gamma$ and small
    type~$C$, when $C'$ is the result of substituting some occurrences
    of~$A$ by~$B$ in~$C$, we have that~$\Gamma \vdash_{\eptt} t:C$ iff~$\Gamma
    \vdash_{\eptt} t:C'$.
  \end{enumerate}
\end{thm}

\begin{thm}\label{thm-duality-eptt}
  In the universe~$\uzero$ of~$\eptt$:
  \begin{enumerate}[label=(\roman*)]
    \item $\op{\phantom{A}}$ is an involution.
    \item $\times$ and~$+$ are dual type constructors under~$\op{\phantom{A}}$.
    \item $\to$ and~$\cofunc$ are dual type constructors under~$\op{\phantom{A}}$.
    \item $\Pi$ and~$\Sigma$ are dual type constructors under~$\op{\phantom{A}}$.
  \end{enumerate}
\end{thm}

Based on Theorem~\ref{thm-duality-eptt}, we shall add to~$\eptt$ the equality
rules in Table~\ref{tab-eq-rules-opp}.

\begin{table}[!ht]
  \centering
  \begin{tabular}{@{}|c|c|@{}}
    \hline
    \rule{0pt}{1.8em} \pttopfunceq{} & \pttopcofunceq{} \rule[-1.25em]{0pt}{3em}
    \\
    \hline
    \rule{0pt}{1.8em} \pttopprodeq{} & \pttopdijueq{} \rule[-1.25em]{0pt}{3em}
    \\
    \hline
    \rule{0pt}{1.8em} \pttoppieq{} & \pttopsigeq{} \rule[-1.25em]{0pt}{3em}
    \\
    \hline
    \rule{0pt}{1.8em} \pttopopeq{} & \rule[-1.25em]{0pt}{3em}
    \\
    \hline
  \end{tabular}
  \caption{Equality rules for the opposite type
  constructor.}\label{tab-eq-rules-opp}
\end{table}

As in~$\eptt$ we do not have basic types,\footnote{By \emph{basic types} we
mean types without type constructors.}  we suppose that
types of~$\eptt$ are generated by a denumerable set of type
variables~$V = \{\alpha, \alpha_1, \ldots, \beta, \beta_1, \ldots\}$ which
represent arbitrary (possibly dependent) basic types. In types~$\pit{x}{A}{B}$
and~$\sigt{x}{A}{B}$ we shall call~$A$ the \emph{generating type}, considering
that~$B(x)$ is a family of types generated on~$A$.

\begin{defn}
  Let~$A$ be a type of~$\ptt$. The \emph{dual} of~$A$, which
  will be denoted by~$\dual{A}$, is the result of exchange~$\to$
  and~$\cofunc$ and swap the type on the left-hand side with the type on the
  right-hand side of such constructors, exchange~$\times$ and~$+$,
  exchange~$\Pi$ and~$\Sigma$, and exchange each type variable~$\alpha$
  by~$\op{\alpha}$, letting the generating types unchanged.
\end{defn}

\begin{thm}[Principle of duality in~$\eptt$]
  Let~$A$ be a small type of~$\eptt$, then $\vdash_{\eptt} A = \op{\dual{A}} :
  \uzero$.
\end{thm}

As it was pointed out in~\cite{agudelo-sicard-2021}, in~$\ptt$ the
type constructor~$\to$ can be defined by~$A \to B \eqdef \Pi x:A.B$,
when~$x$ is not free in~$B$, as in~$\itt$ (because
$A \to B \equiv_s \Pi x:A.B$, when~$x$ is not free in~$B$). However,
while in~$\itt$ the type constructor $\times$ can be defined by
$A \times B \eqdef \sigt{x}{A}{B}$, when~$x$ is not free in~$B$,
in~$\ptt$ this definition is not possible (because
$\op{A \times B} \nequivt{\uzero} \op{\sigt{x}{A}{B}}$). The same happens
in~$\eptt$. However, in~$\eptt$ we have that
$B \cofunc A \equiv_s \sigt{x}{\op{A}}{B}$, when~$x$ is not free
in~$B$; which allows us to define the type constructor~$\cofunc$ by
$B \cofunc A \eqdef \sigt{x}{\op{A}}{B}$, when~$x$ is not free in~$B$. This
shows that co-function types in~$\eptt$ can be viewed as a kind of product,
different of the Cartesian product $\times$.\footnote{The Cartesian product
in~$\itt$ is defined as~$A \times B \equiv_s \sigt{x}{A}{B}$, when~$x$ is not
free in~$B$. This definition corresponds to~$B \cofunc \op{A}$
in~$\eptt$, because $B \cofunc \op{A} \eqdef \sigt{x}{\op{\op{A}}}{B} \equiv_s
\sigt{x}{A}{B}$.} While the opposite of a co-function type is a function type,
the opposite of a Cartesian product is a disjoint union.

Moreover, as it was also pointed out in~\cite{agudelo-sicard-2021},
while in~$\itt$ the type constructors~$+$ and~$\Sigma$ cannot be
defined by means of the other type constructors, in~$\ptt$ the type
constructor~$+$ can be defined
by~$A + B \eqdef \op{\op{A} \times \op{B}}$ (because
$A + B \sequivt{\uzero} \op{\op{A} \times \op{B}}$) and~$\Sigma$ can be
defined by~$\sigt{x}{A}{B} \eqdef \op{\pit{x}{A}{\op{B}}}$ (because
$\sigt{x}{A}{B} \sequivt{\uzero} \op{\pit{x}{A}{\op{B}}}$). These definitions
are also valid in~$\eptt$.

Taking into account the possible definitions of type constructors
described in the previous paragraphs, the sets of constructors
$\big\{ \Pi, \times, \op{\phantom{A}} \big\}$,
$\big\{ \Pi, +, \op{\phantom{A}} \big\}$,
$\big\{ \Sigma, \times, \op{\phantom{A}} \big\}$ and
$\big\{ \Sigma, +, \op{\phantom{A}} \big\}$ are complete for~$\eptt$.

Duality in~$\eptt$, and the equality rules in
Table~\ref{tab-eq-rules-opp}, also allows us to carry the
opposite type constructors to basic types as stated below.

\begin{defn}
  A type~$A$ of~$\eptt$ is in \emph{opposite normal form} if the
  opposite constructor is only applied to type variables in~$A$.
\end{defn}

\begin{thm}\label{thm-opp-normal-form}
  Every small type of~$\eptt$ is equal to a type in opposite normal form.
\end{thm}

\section{Intuitive interpretations of opposite and co-function
types}\label{sec-opp-cof-int}

Martin-Löf~\cite[p.~5]{martin-lof-1984} provides four different
intuitive interpretations of judgements in intuitionistic type
theory. In one of such interpretations, types are understood as
\enquote{intentions}. However, as the notion of intention is too
vague, we shall not consider this interpretation here. The other three
interpretations are shown in Table~\ref{tab-interp-jud-itt}.

\begin{table}[!ht]
  \center
  \begin{tabular}{@{}p{3cm}p{4.5cm}p{3cm}@{}}
    \toprule
    $\mathbf{A : Set}$ & $\mathbf{a : A}$ & \textbf{Inhabitation of
    $\mathbf{A}$}
    \\
    \hline
    $A$ is a proposition & $a$ is a proof (construction) of
                           proposition $A$ & $A$ is true
    \\[1.75em]
    $A$ is a problem (task) & $a$ is a method of solving the problem (doing the
                              task) $A$ & $A$ is solvable
    \\[1.75em]
    $A$ is a set & $a$ is an element of the set $A$ & $A$ is non-empty
    \\
    \bottomrule
  \end{tabular}
  \caption{Interpretations of Martin-Löf's judgements
    forms.}\label{tab-interp-jud-itt}
\end{table}

The first interpretation corresponds to the so-called
\emph{propositions-as-types interpretation}. With respect to the second
interpretation, Martin-Löf explains that:

\begin{displayquote}
  {}[This intepretation] is very close to programming,
  \enquote{$a$ is a method ...} can be read as \enquote{$a$~is a
    program \ldots}. Since programming languages have a formal
  notation for the program~$a$, but not for~$A$, we complete the
  sentence with \enquote{\ldots which meets the specification~$A$}. In
  Kolmogorov's interpretation, the word problem refers to something to
  be done and the word program to how to do it.
\end{displayquote}

In the third interpretation types are interpreted as sets.

Under each one of the interpretations of types, constructors
$\to, \times, +, \Pi, \Sigma$ have their respective interpretations
in~$\itt$, which are shown in Table~\ref{tab-interp-const-itt}. In
such table, symbols $\impl, \conj, \disj, \forall, \exists$ are the
intuitionistic logical constants for implication, conjunction,
disjunction, universal quantification and existential quantification,
respectively. Moreover, a many-sorted version of first-order intuitionistic
logic must be considered~\cite{agudelo-sicard-2021}, and the
sort of variables are indicated by superscripts. In the third column,
by~$\{B_x\}_{x \in A}$ we denote a \emph{family of problems (or
  problem specifications)} that is parametrised by elements in~$A$;
that is, for each~$x \in A$, we have that~$B_x$ is a specification of
a problem. By a \emph{general method for solving a family of
  problems~$\{B_x\}_{x \in A}$} we mean a single method that for each
$x \in A$ gives a solution for~$B_x$, and by a \emph{particular method
  for solving a problem in a family~$\{B_x\}_{x \in A}$} we mean a
method that for some $x \in A$ gives a solution for~$B_x$.

\begin{table}[!ht]
  \center
  \begin{tabular}{@{}cp{3cm}p{5cm}p{4.7cm}@{}}
    \toprule
    & \textbf{Types interpreted as propositions} & \textbf{Types
    interpreted as problems} & \textbf{Types interpreted as sets}
    \\
    \hline
    $A \to B$ & \centering $A \impl B$ & Methods that transforms any solution
    of~$A$ into a solution of~$B$ & Set of functions from~$A$ to~$B$
    \\[0.5em]
    $A \times B$ & \centering $A \conj B$ & Methods that solve~$A$ and~$B$ &
    Cartesian product of~$A$ and~$B$
    \\[0.5em]
    $A + B$ & \centering $A \disj B$ & Methods that solve~$A$ or~$B$  &
    Disjoint union of~$A$ and~$B$
    \\[0.5em]
    $\pit{x}{A}{B}$ & \centering $\foralls{x}{A}{B}$ & General methods for
    solving the family of problems $\{B_x\}_{x\in A}$ & Generalised Cartesian
    product of the family of sets
    $\{B_x\}_{x\in A}$
    \\[0.5em]
    $\sigt{x}{A}{B}$ & \centering $\existss{x}{A}{B}$ & Particular methods for
    solving a problem in the family $\{B_x\}_{x\in A}$ & Generalised disjoint
    union of the family of sets $\{B_x\}_{x\in A}$
    \\
    \bottomrule
  \end{tabular}
  \caption{Interpretation of constructors
  in~$\itt$.}\label{tab-interp-const-itt}
\end{table}

In~$\eptt$, the type constructors~$\to, \times, +, \Pi, \Sigma$ can be
interpreted in the same way that in~$\itt$, under every interpretation of
types. In the logical interpretation, logical constants correspond to those
of~$\epls$. Now, we shall explain how opposite types and co-function types can
be interpreted in every interpretation of types.

\begin{enumerate}[label=(\roman*),leftmargin=3ex]

\item For the interpretation of types as propositions, as we mentioned
  in Section~\ref{sec-ptt-to-eptt}, the propositions-as-types
  correspondence between~$\pls$ and~$\ptt$ can be extended to obtain a
  propositions-as-types correspondence between~$\epls$
  and~$\eptt$. Under such extended correspondence:
  \begin{itemize}[leftmargin=2ex]
  \item The opposite type constructor corresponds to the negation
    of~$\epls$.  Thus, following the same interpretation of negation
    in~$\pls$, the negation in~$\epls$ represents falsity and
    consequently the inhabitation of~$\op{A}$ can be understood as
    \enquote{$\sneg A$ is true} or as \enquote{$A$ is false}, and
    $a : \op{A}$ can be understood as \enquote{$a$ is a proof of the
      negation of~$A$} or as \enquote{$a$ is a refutation
      of~$A$}. Under this interpretations, the dualities in
    Section~\ref{sec-dual-eptt} make sense.

  \item The co-function type constructor corresponds to co-implication
    in~$\epls$.
  \end{itemize}

\item Under the interpretation of types as problems:
  \begin{itemize}[leftmargin=2ex]
  \item A type $\op{A}$ can be understood as a specification of a
    problem whose solution excludes a solution of~$A$. Consequently,
    inhabitation of~$\op{A}$ can be understood as \enquote{$\op{A}$ is
      solvable} or as \enquote{$A$ is unsolvable}, and $a: \op{A}$ can be
    understood as \enquote{$a$ is a method that solves $\op{A}$} or as
    \enquote{$a$ is a method that shows the unsolvability of
      $A$}. Under this interpretations, the dualities in
    Section~\ref{sec-dual-eptt} make sense.

  \item As~$B \cofunc A \equivt{\uzero} \op{A} \times B$, thus $c:B \cofunc A$
    can be understood as \enquote{$c$~is a method that solves $\op{A}$ (or
    shows the unsolvability of~$A$) and that solves~$B$}.
  \end{itemize}

\item Under the interpretation of types as sets:
  \begin{itemize}[leftmargin=2ex]
  \item It is harder to glimpse an intuitive interpretation for
    opposite types. Taking into consideration that, if we interpret
    types as sets, the inhabitation of type~$A$ means that~$A$ is
    non-empty, we might initially think that the duality in this case
    is between non-emptiness and emptiness.  Accordingly, the
    inhabitation of~$\op{A}$ should be interpreted as
    \enquote{$\op{A}$ is non-empty} or as \enquote{$A$ is empty}. But
    it does not make sense that the non-emptiness of a
    set~(i.e.~$\op{A}$) leads to the emptiness of another
    set~(i.e.~$A$). However, considering an intentional conception of
    sets, under which it can be roughly said that a set consists in a
    collection of individuals that fall under an associated
    concept,\footnote{For an historical and philosophical analysis of
      the concept of set, the differentiation between intensional and
      extensional, and also between essential and accidental
      conceptions, see~\cite[Chapter III]{ganstrom-2011}.}, which we
    shall denote by~$\conc{A}$, we can understood~$\op{A}$ as the
    collection of individuals that fall under the \enquote{dual
      concept} to~$\conc{A}$. Consequently, the inhabitation
    of~$\op{A}$ must be understood as \enquote{there are individuals
      that fall under the concept~$\conc{\op{A}}$} or as
    \enquote{there are individuals that fall under the dual concept
      to~$\conc{A}$}. Under this (somewhat blurred) intentional
    interpretation of sets, the dualities in
    Section~\ref{sec-dual-eptt} make sense. However, taking into
    account Theorem~\ref{thm-opp-normal-form}, for a full
    interpretation of types as sets we just need to have the
    interpretation of opposite basic types, but in this endeavour it
    is necessary to define what exactly means \enquote{dual concept}
    (at least for concepts associated with basic types), which is not
    a simple task and is left for future work.

  \item As~$B \cofunc A \equivt{\uzero} \op{A} \times B$, thus $c:B \cofunc A$
    can be understood as \enquote{$c$ is a pair of individuals, the
      first one falling under the concept~$\conc{\op{A}}$ (or under the concept
      dual to~$\conc{A}$) and the second one falling under the
      concept~$\conc{B}$}.
  \end{itemize}
\end{enumerate}

\section*{Acknowledgements}
\pdfbookmark[1]{Acknowledgements}{sec-acknowledgements}

The authors are deeply grateful to the anonymous reviewers for their valuable
comments and suggestions which led to important improvements to the article.

\addcontentsline{toc}{section}{\refname}

\bibliographystyle{eptcs}
\bibliography{ptt-opp-dual}
\end{document}

%% file: commands.tex
\usepackage{amstext, amssymb, amsthm, amsmath, tikz,
bussproofs,stmaryrd}

\newcommand{\msf}[1]{\mathsf{#1}}

\newcommand{\mr}[1]{\mathrm{#1}}

\newcommand{\eqdef}{\stackrel{\text{\tiny \upshape def}}{=}}

\newcommand{\conj}{\wedge}
\newcommand{\disj}{\vee}
\newcommand{\impl}{\supset}
\newcommand{\coimpl}{\prec}
\newcommand{\impli}{\rightarrow}
\newcommand{\coimpli}{\Yleft}

\newcommand{\sneg}{\sim} 

\newcommand{\foralls}[3]{(\forall #1^{#2})#3}
\newcommand{\existss}[3]{(\exists #1^{#2})#3}

\newcommand{\blc}{\mathrel{\#}}

\newcommand{\equivt}[1]{\equiv_{#1}}

\newcommand{\nequivt}[1]{\not \equiv_{#1}}

\newcommand{\sequivt}[1]{\mathrel{\equiv_{#1}^{\msf{s}}}}

\newcommand{\nsequivt}[1]{\mathrel{{\not \equiv}_{#1}^\msf{s}}}

\newcommand{\abst}[2]{(\lambda #1)#2} 
\newcommand{\apl}[2]{\msf{Ap}(#1,#2)} 
\newcommand{\pair}[2]{(#1,#2)} 

\newcommand{\projl}{\msf{p}}
\newcommand{\projr}{\msf{q}}
\newcommand{\proj}[2]{
  \ifnum1=#1
    \projl(#2)
  \else
    \projr(#2)
  \fi
}

\newcommand{\injl}{\msf{i}}
\newcommand{\injr}{\msf{j}}
\newcommand{\inj}[2]{
  \ifnum1=#1
    \injl(#2)
  \else
    \injr(#2)
  \fi
}
\newcommand{\eltd}[5]{\msf{D}(#1,(#2)#3(#2),(#4)#5(#4))} 
\newcommand{\elts}[4]{\msf{E}(#1,(#2,#3)#4(#2,#3))} 

\newcommand{\pit}[3]{(\Pi #1:#2)#3} 
\newcommand{\sigt}[3]{(\Sigma #1:#2)#3} 
\newcommand{\cofunc}{\mathbin{\Yleft}} 

\newcommand{\op}[1]{\overline{#1}}

\newcommand{\dual}[1]{#1^{\star}}

\newcommand{\uset}{\mathtt{Set}}

\newcommand{\uzero}{\mr{U_0}}
\newcommand{\uone}{\mr{U_1}}
\newcommand{\univ}{\mr{U}}

\newcommand{\nlcoimpli}[1]{%
  \AxiomC{$\sneg A$}%
  \AxiomC{$B$}%
  \RightLabel{#1}%
  \BinaryInfC{$B \coimpl A$}%
  \DisplayProof%
}

\newcommand{\nlcoimplel}[1]{%
  \AxiomC{$B \coimpl A$}%
  \RightLabel{#1}%
  \UnaryInfC{$\sneg A$}%
  \DisplayProof%
}

\newcommand{\nlcoimpler}[1]{%
  \AxiomC{$B \coimpl A$}%
  \RightLabel{#1}%
  \UnaryInfC{$B$}%
  \DisplayProof%
}

\newcommand{\nlncoimpli}[1]{
  \AxiomC{$[\sneg A]$}%
  \noLine%
  \UnaryInfC{$\vdots$}%
  \noLine%
  \UnaryInfC{}%
  \noLine%
  \UnaryInfC{$\sneg B$}%
  \RightLabel{#1}%
  \UnaryInfC{$\sneg (B \coimpl A)$}%
  \DisplayProof%
}

\newcommand{\nlncoimple}[1]{%
  \AxiomC{$\sneg A$}%
  \AxiomC{$\sneg(B \coimpl A)$}%
  \RightLabel{#1}%
  \BinaryInfC{$\sneg B$}%
  \DisplayProof%
}

\newcommand{\pttopf}[1]{%
  \AxiomC{$A : \uset$}%
  \RightLabel{#1}%
  \UnaryInfC{$\op{A} : \uset$}%
  \DisplayProof%
}

\newcommand{\pttopopi}{%
  \AxiomC{$a : A$}%
  \UnaryInfC{$a : \op{\op{A}}$}%
  \DisplayProof%
}

\newcommand{\pttopope}{%
  \AxiomC{$a : \op{\op{A}}$}%
  \UnaryInfC{$a : A$}%
  \DisplayProof%
}

\newcommand{\pttopfunci}{%
  \AxiomC{$a : A$}%
  \AxiomC{$b : \op{B}$}%
  \BinaryInfC{$\pair{a}{b} : \op{A \to B}$}%
  \DisplayProof%
}

\newcommand{\pttopfuncel}{%
  \AxiomC{$c : \op{A \to B}$}%
  \UnaryInfC{$\proj{1}{c} : A$}%
  \DisplayProof%
}

\newcommand{\pttopfuncer}{%
  \AxiomC{$c : \op{A \to B}$}%
  \UnaryInfC{$\proj{2}{c} : \op{B}$}%
  \DisplayProof%
}

\newcommand{\pttopprodil}{%
  \AxiomC{$a : \op{A}$}%
  \UnaryInfC{$\inj{1}{a} : \op{A \times B}$}%
  \DisplayProof%
}

\newcommand{\pttopprodir}{%
  \AxiomC{$b : \op{B}$}%
  \UnaryInfC{$\inj{2}{b} : \op{A \times B}$}%
  \DisplayProof%
}

\newcommand{\pttopprode}{%
  \def\defaultHypSeparation{\hskip 0.25cm}%
  \AxiomC{$c : \op{A \times B}$}%
  \AxiomC{$(x : \op{A})$}%
  \noLine%
  \UnaryInfC{$\vdots$}%
  \noLine%
  \UnaryInfC{}%
  \noLine%
  \UnaryInfC{$d(x) : C(\inj{1}{x})$}%
  \AxiomC{$(y : \op{B})$}%
  \noLine%
  \UnaryInfC{$\vdots$}%
  \noLine%
  \UnaryInfC{}%
  \noLine%
  \UnaryInfC{$e(y) : C(\inj{2}{y})$}%
  \TrinaryInfC{$\eltd{c}{x}{d}{y}{e} : C(c)$}%
  \DisplayProof%
}

\newcommand{\pttopdijui}{%
  \AxiomC{$a : \op{A}$}%
  \AxiomC{$b : \op{B}$}%
  \BinaryInfC{$\pair{a}{b} : \op{A + B}$}%
  \DisplayProof%
}

\newcommand{\pttopdijuel}{%
  \AxiomC{$c : \op{A + B}$}%
  \UnaryInfC{$\proj{1}{c} : \op{A}$}%
  \DisplayProof%
}

\newcommand{\pttopdijuer}{%
  \AxiomC{$c : \op{A + B}$}%
  \UnaryInfC{$\proj{2}{c} : \op{B}$}%
  \DisplayProof%
}

\newcommand{\pttoppii}{%
  \AxiomC{$a : A$}%
  \AxiomC{$b : \op{B(a)}$}%
  \BinaryInfC{$\pair{a}{b} : \op{\pit{x}{A}{B(x)}}$}%
  \DisplayProof%
}

\newcommand{\pttoppie}{%
  \AxiomC{$c : \op{\pit{x}{A}{B(x)}}$}%
  \AxiomC{$(x: A, y : \op{B(x)})$}%
  \noLine%
  \UnaryInfC{$\vdots$}%
  \noLine%
  \UnaryInfC{}%
  \noLine%
  \UnaryInfC{$d(x,y) : C((x,y))$}%
  \BinaryInfC{$\elts{c}{x}{y}{d} : C(c)$}%
  \DisplayProof%
}

\newcommand{\pttopsigi}{%
  \AxiomC{$(x : A)$}%
  \noLine%
  \UnaryInfC{$\vdots$}%
  \noLine%
  \UnaryInfC{}%
  \noLine%
  \UnaryInfC{$b(x) : \op{B(x)}$}%
  \UnaryInfC{$\abst{x}{b(x)} : \op{\sigt{x}{A}{B(x)}}$}%
  \DisplayProof%
}

\newcommand{\pttopsige}{%
  \AxiomC{$b : \op{\sigt{x}{A}{B(x)}}$}%
  \AxiomC{$a : A$}%
  \BinaryInfC{$\apl{b}{a} : \op{B(a)}$}%
  \DisplayProof%
}

\newcommand{\pttcofuncf}[1]{%
  \AxiomC{$A : \uset$}%
  \AxiomC{$B : \uset$}%
  \RightLabel{#1}%
  \BinaryInfC{$B \cofunc A : \uset$}%
  \DisplayProof%
}

\newcommand{\pttcofunci}[1]{%
  \AxiomC{$a : \op{A}$}%
  \AxiomC{$b : B$}%
  \RightLabel{#1}%
  \BinaryInfC{$\pair{a}{b} : B \cofunc A$}%
  \DisplayProof%
}

\newcommand{\pttcofuncel}[1]{%
  \AxiomC{$c : B \cofunc A$}%
  \RightLabel{#1}%
  \UnaryInfC{$\proj{1}{c} : \op{A}$}%
  \DisplayProof%
}

\newcommand{\pttcofuncer}[1]{%
  \AxiomC{$c : B \cofunc A$}%
  \RightLabel{#1}%
  \UnaryInfC{$\proj{2}{c} : B$}%
  \DisplayProof%
}

\newcommand{\pttopcofunci}[1]{%
  \AxiomC{$(x : \op{A})$}%
  \noLine%
  \UnaryInfC{$\vdots$}%
  \noLine%
  \UnaryInfC{}%
  \noLine%
  \UnaryInfC{$b(x) : \op{B}$}%
  \RightLabel{#1}%
  \UnaryInfC{$\abst{x}{b(x)} : \op{B \cofunc A}$}%
  \DisplayProof%
}

\newcommand{\pttopcofunce}[1]{%
  \AxiomC{$c : \op{B \cofunc A}$}%
  \AxiomC{$a : \op{A}$}%
  \RightLabel{#1}%
  \BinaryInfC{$\apl{c}{a} : \op{B}$}%
  \DisplayProof%
}

\newcommand{\pttopfunceq}[1]{%
  \AxiomC{$A : \uzero$}%
  \AxiomC{$B : \uzero$}%
  \RightLabel{#1}%
  \BinaryInfC{$\op{A \to B} = \op{B} \cofunc \op{A} : \uzero$}%
  \DisplayProof%
}

\newcommand{\pttopcofunceq}[1]{%
  \AxiomC{$A : \uzero$}%
  \AxiomC{$B : \uzero$}%
  \RightLabel{#1}%
  \BinaryInfC{$\op{B \cofunc A} = \op{A} \to \op{B} : \uzero$}%
  \DisplayProof%
}

\newcommand{\pttopprodeq}[1]{%
  \AxiomC{$A : \uzero$}%
  \AxiomC{$B : \uzero$}%
  \RightLabel{#1}%
  \BinaryInfC{$\op{A \times B} = \op{A} + \op{B} : \uzero$}%
  \DisplayProof%
}

\newcommand{\pttopdijueq}[1]{%
  \AxiomC{$A : \uzero$}%
  \AxiomC{$B : \uzero$}%
  \RightLabel{#1}%
  \BinaryInfC{$\op{A + B} = \op{A} \times \op{B} : \uzero$}%
  \DisplayProof%
}

\newcommand{\pttoppieq}[1]{%
  \AxiomC{$A : \uzero$}%
  \AxiomC{$B : \uzero$}%
  \RightLabel{#1}%
  \BinaryInfC{$\op{\pit{x}{A}{B}} = \sigt{x}{A}{\op{B}} : \uzero$}%
  \DisplayProof%
}

\newcommand{\pttopsigeq}[1]{%
  \AxiomC{$A : \uzero$}%
  \AxiomC{$B : \uzero$}%
  \RightLabel{#1}%
  \BinaryInfC{$\op{\sigt{x}{A}{B}} = \pit{x}{A}{\op{B}} : \uzero$}%
  \DisplayProof%
}

\newcommand{\pttopopeq}[1]{%
  \AxiomC{$A : \uzero$}%
  \RightLabel{#1}%
  \UnaryInfC{$\op{\op{A}} = A : \uzero$}%
  \DisplayProof%
}

\newcommand{\pttetaconvabst}[2]{%
  \AxiomC{$t : #1$}%
  \RightLabel{#2}%
  \UnaryInfC{$\abst{x}{\apl{t}{x}} = t : #1$}%
  \DisplayProof%
}

\newcommand{\pttetaconvpair}[2]{%
  \AxiomC{$t : #1$}%
  \RightLabel{#2}%
  \UnaryInfC{$\pair{\proj{1}{t}}{\proj{2}{t}} = t : #1$}%
  \DisplayProof%
}

\newcommand{\pttcoetaconvseld}[2]{%
  \AxiomC{$t : #1$}%
  \RightLabel{#2}%
  \UnaryInfC{$\eltd{t}{x}{\injl}{y}{\injr} = t : #1$}%
  \DisplayProof%
}

\newcommand{\pttcoetaconvsels}[2]{%
  \AxiomC{$t : #1$}%
  \RightLabel{#2}%
  \UnaryInfC{$\elts{t}{x}{y}{} = t : #1$}%
  \DisplayProof%
}

\newcommand{\ndtilimpli}[1]{%
  \AxiomC{$[A]$}%
  \noLine%
  \UnaryInfC{$\vdots$}%
  \noLine%
  \UnaryInfC{}%
  \UnaryInfC{$B$}%
  \RightLabel{#1}%
  \UnaryInfC{$A \impl B$}%
  \DisplayProof%
}

\newcommand{\ndtilimplid}[1]{%
  \AxiomC{$\cass{A}$}%
  \noLine%
  \UnaryInfC{$\vdots$}%
  \noLine%
  \UnaryInfC{}%
  \refutationLine%
  \UnaryInfC{$B$}%
  \RightLabel{#1}%
  \refutationLine%
  \RightLabel{#1}%
  \UnaryInfC{$B \coimpl A$}%
  \DisplayProof%
}

\newcommand{\ndtilimple}[1]{%
  \AxiomC{}%
  \UnaryInfC{$A$}%
  \AxiomC{}%
  \UnaryInfC{$A \impl B$}%
  \RightLabel{#1}%
  \BinaryInfC{$B$}%
  \DisplayProof%
}

\newcommand{\ndtilimpled}[1]{%
  \AxiomC{}%
  \refutationLine%
  \UnaryInfC{$A$}%
  \AxiomC{}%
  \refutationLine%
  \UnaryInfC{$B \coimpl A$}%
  \refutationLine%
  \RightLabel{#1}%
  \BinaryInfC{$B$}%
  \DisplayProof%
}

\newcommand{\ndtilnimpli}[1]{%
  \AxiomC{}%
  \UnaryInfC{$A$}%
  \AxiomC{}%
  \refutationLine%
  \UnaryInfC{$B$}%
  \refutationLine%
  \RightLabel{#1}%
  \BinaryInfC{$A \impl B$}%
  \DisplayProof%
}

\newcommand{\ndtilnimplel}[1]{%
  \AxiomC{}%
  \refutationLine%
  \UnaryInfC{$A \impl B$}%
  \RightLabel{#1}%
  \UnaryInfC{$A$}%
  \DisplayProof%
}

\newcommand{\ndtilnimpler}[1]{%
  \AxiomC{}%
  \refutationLine%
  \UnaryInfC{$A \impl B$}%
  \refutationLine%
  \RightLabel{#1}%
  \UnaryInfC{$B$}%
  \DisplayProof%
}

\newcommand{\ndtilncoimpli}[1]{%
  \AxiomC{}%
  \refutationLine%
  \UnaryInfC{$A$}%
  \AxiomC{}%
  \UnaryInfC{$B$}%
  \RightLabel{#1}%
  \BinaryInfC{$B \coimpl A$}%
  \DisplayProof%
}

\newcommand{\ndtilncoimplel}[1]{%
  \AxiomC{}%
  \UnaryInfC{$B \coimpl A$}%
  \RightLabel{#1}%
  \UnaryInfC{$B$}%
  \DisplayProof%
}

\newcommand{\ndtilncoimpler}[1]{%
  \AxiomC{}%
  \UnaryInfC{$B \coimpl A$}%
  \refutationLine%
  \RightLabel{#1}%
  \UnaryInfC{$A$}%
  \DisplayProof%
}